\documentclass[sigconf]{acmart}
\AtBeginDocument{%
  }

\copyrightyear{2026}
\acmYear{2026}
\setcopyright{cc}
\setcctype{by}
\acmConference[WWW '26] {Proceedings of the ACM Web Conference 2026}{April 13--17, 2026}{Dubai, United Arab Emirates.}
\acmBooktitle{Proceedings of the ACM Web Conference 2026 (WWW '26), April 13--17, 2026, Dubai, United Arab Emirates}
\acmISBN{979-8-4007-2307-0/2026/04}
\acmDOI{10.1145/3774904.3792830}

\usepackage{balance}
\usepackage{breqn}
\usepackage{algorithm} 
\usepackage{algorithmic}
\usepackage{multirow} 
\usepackage{makecell}

\begin{document}

\title{A Long-term Value Prediction Framework In Video Ranking}

\author{Huabin Chen}
\affiliation{%
\institution{Alibaba Group}
  \city{Hangzhou}
  \country{China}}
\email{chenhuabin.chb@alibaba-inc.com}

\author{Xinao Wang}
\affiliation{%
  \institution{Alibaba Group}
  \city{Hangzhou}
  \country{China}}
\email{wangxinao.wxa@alibaba-inc.com}

\author{Huiping Chu}
\affiliation{%
  \institution{Alibaba Group}
  \city{Hangzhou}
  \country{China}}
\email{yueqi.chp@alibaba-inc.com}

\author{Keqin Xu}
\affiliation{%
  \institution{Alibaba Group}
  \city{Hangzhou}
  \country{China}}
\email{xukeqin.xkq@alibaba-inc.com}

\author{Chenhao Zhai}
\affiliation{%
  \institution{Tsinghua University}
  \city{Shenzhen}
  \country{China}}
\email{dch23@mails.tsinghua.edu.cn}

\author{Chenyi Wang}
\affiliation{%
  \institution{Alibaba Group}
  \city{Hangzhou}
  \country{China}}
\email{wangchenyi.wcy@alibaba-inc.com}

\author{Kai Meng}
\affiliation{%
  \institution{Alibaba Group}
  \city{Hangzhou}
  \country{China}}
\email{mengkai.meng@alibaba-inc.com}

\author{Yuning Jiang}
\authornote{Corresponding author}
\affiliation{%
  \institution{Alibaba Group}
  \city{Hangzhou}
  \country{China}}
\email{mengzhu.jyn@alibaba-inc.com}

\renewcommand{\shortauthors}{Huabin Chen et al.}

\begin{abstract}

Accurately modeling long-term value (LTV) at the ranking stage of short-video recommendation systems remains a practical challenge. Though production systems and recent research have begun exploring delayed feedback and extended user engagement, modeling LTV with fine-grained attribution and robust positional normalization for billion-scale platforms is underdeveloped. In this work, we present a practical ranking-stage LTV framework that systematically addresses three core challenges: position bias, attribution ambiguity, and temporal limitations.

First, to address position bias in sequential video feeds, we introduce a Position-aware Debias Quantile (PDQ) module that normalizes engagement signals using quantile-based distributions, enabling position-robust LTV estimation without requiring architectural changes. Second, we propose a multi-dimensional attribution module that learns continuous strengths across contextual, behavioral, and content-related signals, moving beyond static rule sets to capture nuanced influences among videos. Explicit noise filtering is incorporated via a customized hybrid loss, improving causal clarity in LTV attribution. Third, our cross-temporal author modeling module constructs censoring-aware, day-level long-term value targets, capturing creator-driven re-engagement over extended time windows. While our framework currently focuses on the author dimension, it is readily extensible to further aspects such as topics or styles.

Extensive offline experiments and online A/B tests demonstrate statistically significant gains in LTV‑related metrics and stable trade‑offs with short‑term objectives.
The framework is realized as task augmentation within an existing ranking model, facilitates billion-scale deployment on Taobao's production system with efficient training and serving,  achieving sustained user engagement improvements while remaining compatible with industrial constraints.

\end{abstract}



\maketitle

\section{Introduction}
The rapid proliferation of short‑video platforms (e.g., TikTok, Instagram Reels) has intensified the industry focus on advanced video recommendation technologies\cite{sun2024cread,liu2022neural}. 
The accurate prediction of long‑term value (LTV) for videos has emerged as a critical technological challenge in this domain. Contemporary approaches typically model multiple objectives individually and then integrate the predictions to estimate overall value metrics, 
including watch time and completion rate. While many approaches emphasize immediate value, long‑term effects remain underexplored at the ranking stage, motivating robust LTV modeling\cite{sun2024cread,zhan2022deconfounding,ovaisi2020correcting}. 
In reality, videos often exhibit extended influence cycles, and a substantial portion of value materializes beyond the immediate session.

\begin{figure}[htbp]
  \centering
  \includegraphics[width=\linewidth]{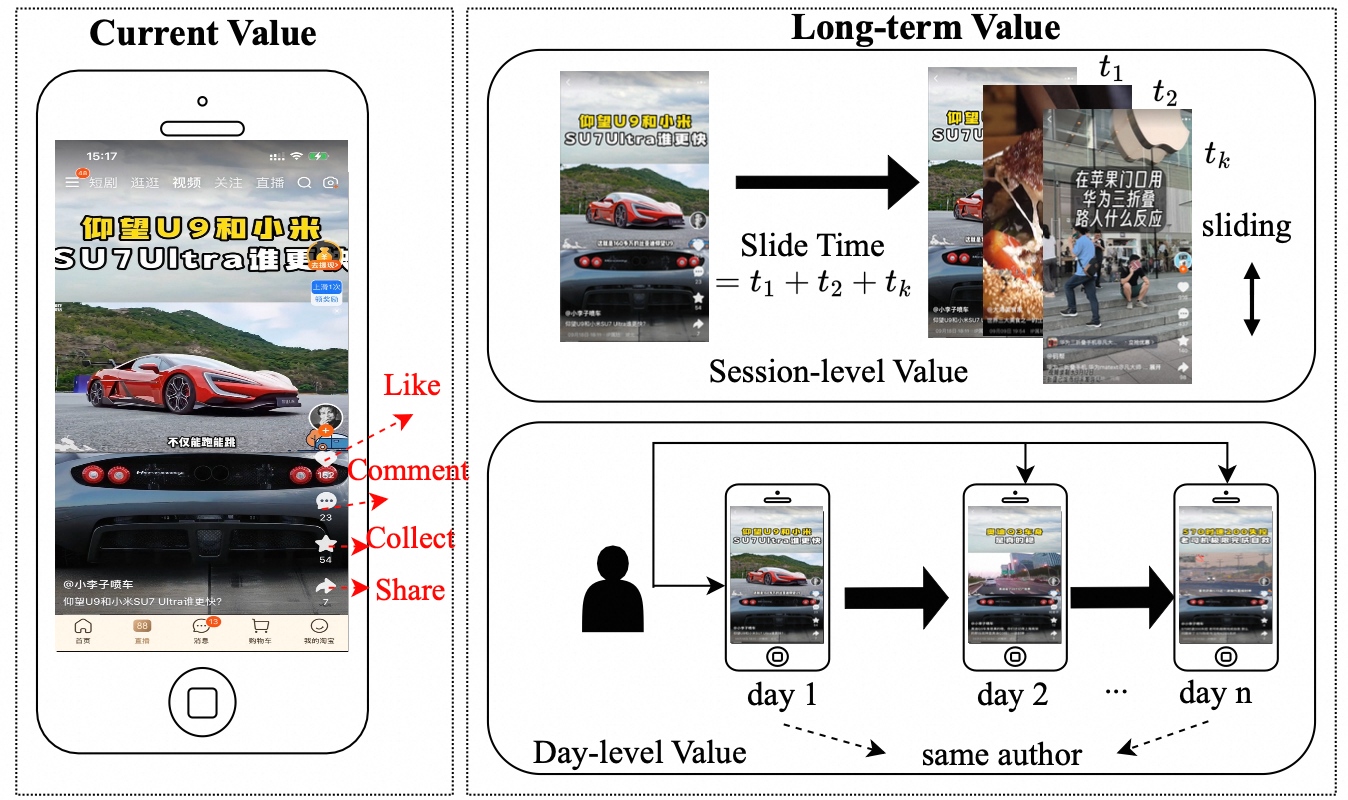}
  \caption{The framework of LTV prediction.}
  \Description{architecture}
  \label{fig:long_term_value_architecture} 
\end{figure}

Page-level value assessment has emerged as a key research direction in multi-stage recommender systems (MRS)\cite{ren2024non,zhao2019deep,liu2022neural}. 
During the re-ranking phase, researchers define sequence-level value through cumulative reward aggregation across constituent videos, calculated using sliding probability metrics. This approach incorporates contextual positionality by considering how a video's value correlates with the positional characteristics of preceding content.
The list-wise frameworks refine this through iterative generator-evaluator architectures\cite{theocharous2015ad}. 
Despite theoretical promise, RL frameworks struggle with computational complexity and training instability in industrial deployments\cite{chen2018stabilizing,taghipour2008hybrid,zheng2018drn,chen2019top,gong2019exact}. 


Despite progress, ranking‑stage LTV remains challenging. We tackle position bias, attribution ambiguity, and limited temporal scope via PDQ (quantile debiasing), multi‑dimensional attribution, and cross‑temporal author modeling.
Position Bias: Direct modeling of raw positional playtime metrics introduces systemic bias, as identical content can exhibit dramatically different slide times depending on the position of exposure.
Attribution Ambiguity: Naive aggregation of subsequent video playtimes fails to establish causal relationships, resulting in noisy value estimates contaminated by irrelevant content.
Temporal Scope Limitation: Intra‑session accumulation overlooks cross‑temporal influence (e.g., creator‑driven re‑engagement) that sustains user value.


To address these limitations, we developed a comprehensive LTV framework for the ranking stage featuring three modeling innovations. As shown in Figure \ref{fig:long_term_value_architecture}, engagement metrics, including follow-up actions, comments, and collections, serve as proxyes for immediate video value assessment. We conceptualize session-level value through cumulative watch time while defining day-level value as cross-temporal engagement patterns manifested in repeated author content consumption. Our framework systematically addresses each identified challenge through the following methods. 

Position Debias Quantile-Based Framework (PDQ):
Inspired by recent advances in debiasing techniques for watch-time prediction \cite{zhan2022deconfounding}, we propose PDQ to address position-related biases in slide time estimation. Our framework employs quantile regression across multipage video interactions to systematically capture the distributional characteristics of user engagement patterns. PDQ quantifies position-dependent temporal dynamics through page-wise quantile analysis, enabling optimal slide time determination within homogeneous page contexts. Unlike existing bias mitigation approaches requiring architectural modifications \cite{ovaisi2020correcting,hofstatter2021mitigating}, our method achieves position bias correction through distributional modeling without altering the underlying model structure. 


Multi‑dimensional Value Attribution: To mitigate attribution inaccuracies, we implement fine‑grained, multi‑dimensional attribution with continuous (learnable) strengths across contextual relevance, behavioral affinity, and content similarity. 
We further refine attribution through author identity, categorical alignment, multimodal similarity, co‑occurrence, and contextual signals, while explicitly filtering unrelated content. 
This hierarchical attribution mechanism enables precise value apportionment through a customized hybrid loss function that mitigates target underestimation. 

Cross‑temporal Author Value Modeling: Adopting a creator‑centric perspective (extensible to topics/styles/memes), we extend session‑level signals with cross‑temporal author modeling, enhancing both dissemination efficiency for high‑quality creators and long‑term engagement. 
Our framework achieves these enhancements through task augmentation within an existing ranking model, eliminating the need for separate re-ranking infrastructure. 
This streamlined implementation facilitates deployment across billion-scale applications while maintaining computational efficiency.

Our contributions are threefold:

\begin{itemize}
\item \textbf{Position‑aware quantile debiasing (PDQ)}: a quantile‑based methodology for mitigating position bias in ranking‑stage LTV prediction without architectural changes, establishing position‑specific normalization baselines.
\item \textbf{Multi‑dimensional attribution}: a hierarchical mechanism that decomposes LTV into contextual, behavioral, and content dimensions with continuous strengths, while filtering irrelevant influences via a hybrid loss.
\item \textbf{Cross-temporal Value Modeling}: 
a ranking‑stage LTV framework that jointly models session‑level and day‑level engagement, validated through comprehensive experiments and deployed at billion-scale industrial applications.
\end{itemize}

\begin{figure}[tp]
  \centering
  \includegraphics[width=1.0\linewidth]{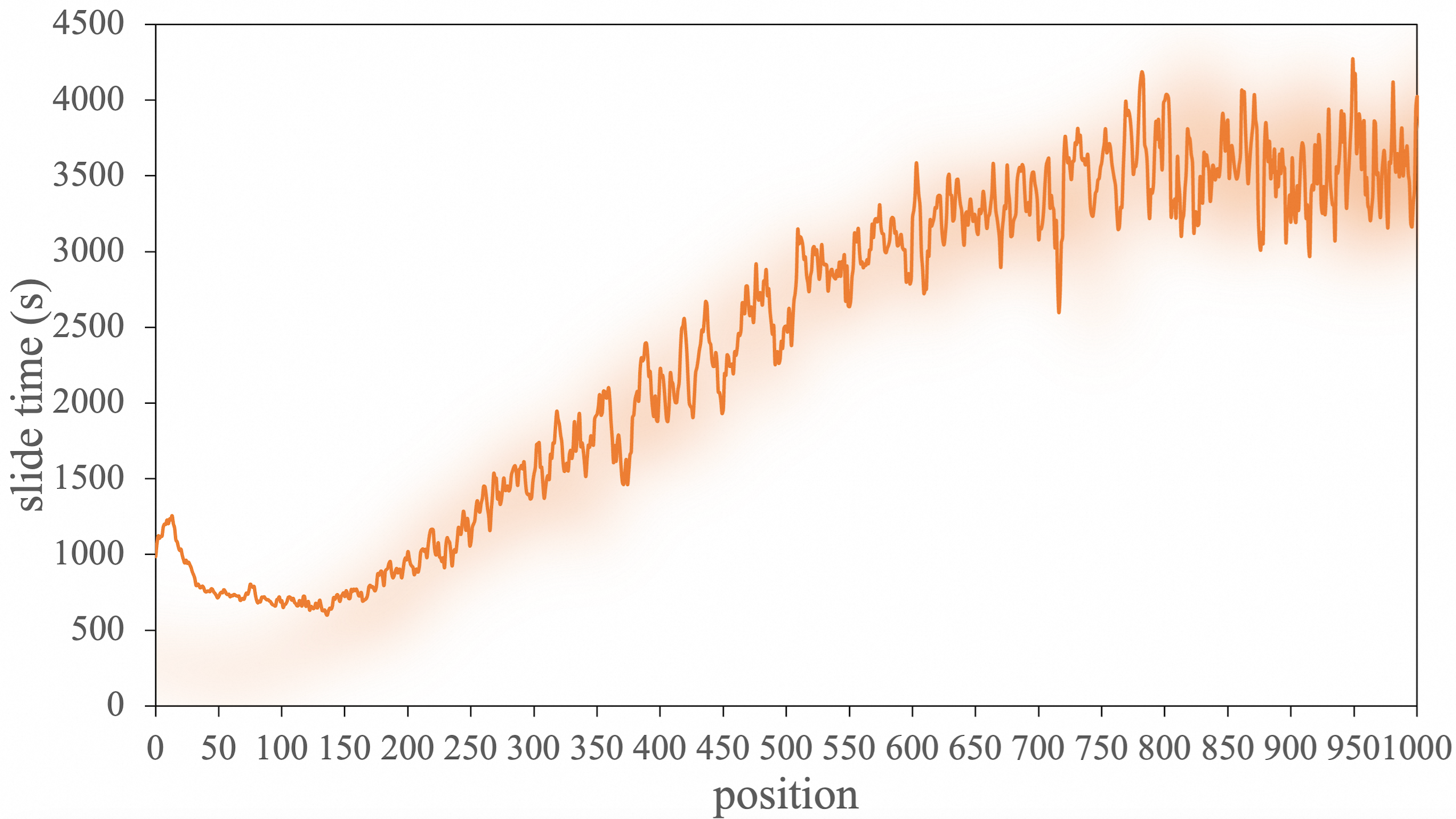}
  \caption{The correlation between slide time and average exposure position over 7 days.}
  \Description{position_bias}
  \label{fig:position_bias}
\end{figure}
\section{Related Works}
This study departs from most previous research by addressing long-term value (LTV) modeling directly within the ranking stage, whereas many existing approaches focus on the re-ranking stage, which operates on a much smaller candidate set after initial ranking.

\subsection{Immediate Response Prediction}

Immediate response prediction has been the dominant paradigm in industrial recommender systems, leveraging deep learning architectures to optimize short-term metrics. Common tasks include click-through rate (CTR) prediction \cite{lin2023map,mao2023finalmlp}, post-click conversion rate (CVR) prediction \cite{ma2018entire,wen2020entire}, and watch time prediction \cite{covington2016deep,zhan2022deconfounding,sun2024cread}, all of which emphasize immediate engagement signals. Watch time prediction estimates the temporal engagement of the user: Covington et al. \cite{covington2016deep} transform direct regression of watch time into inferred probabilities of video clicks, though duration bias can significantly affect this approach. Sun et al. \cite{sun2024cread} introduce a classification–restoration framework to capture ordinal watch-time information, improving the handling of long-tail samples. Zhan et al. \cite{zhan2022deconfounding} apply equal-frequency binning by video duration to mitigate duration bias. Other works have addressed position and selection biases, including regression-based EM algorithms for sparse clicks \cite{wang2018position}, Heckman’s two-stage correction \cite{ovaisi2020correcting}, debiasing for transformer-based reranking \cite{hofstatter2021mitigating}, and propensity estimation without manual judgments \cite{agarwal2019estimating}. 
These works are not very effective in addressing position bias for ranking models and we designed a more effective framework.

\subsection{Long-term Value Prediction}

Most prior work on long-term value (LTV) prediction concentrates on the re‑ranking stage, where listwise context can be explicitly modeled\cite{liu2022neural,xian2019reinforcement,ji2021reinforcement}. Neural re‑ranking typically learns a multivariate listwise scorer to capture cross‑item interactions and positional dependencies, while generator–evaluator paradigms optimize page‑level value under listwise context\cite{ren2024non,theocharous2015ad}. Reinforcement learning(RL) including value‑based, policy‑based, and model‑based variants has been explored for end‑to‑end list optimization and display arrangement\cite{chen2018stabilizing,taghipour2008hybrid,zheng2018drn,chen2019top,bai2019model,zhao2019deep,zhao2018deep}, yet production deployments often face computational overhead and stability issues; moreover, page‑level horizons remain relatively short, limiting sensitivity to longer‑horizon effects. Exploration‑oriented LTV frameworks introduce new metrics and protocols but are orthogonal to efficiency‑centric ranking objectives\cite{su2024long}.
In contrast, ranking-stage LTV must operate over far larger candidate pools under strict latency constraints. This setting favors lightweight, stable techniques that can be integrated without major infrastructure changes. 

A related literature on delayed feedback—primarily in CVR prediction—addresses label bias when outcomes arrive late: Defer ingests real negatives with importance weighting to align observed and true distributions\cite{gu2021real}; ES‑DFM leverages elapsed‑time sampling with instance‑wise importance weights\cite{yang2021capturing}; ULC performs unbiased label correction via auxiliary models with theoretical guarantees\cite{wang2023unbiased}. While these works target CVR rather than video LTV, the underlying challenges (freshness vs label certainty, distribution shift) are analogous. Motivated by these insights, we adopt censoring‑aware day‑level targets and combine them with position‑aware debiasing (PDQ) and multi‑dimensional attribution at the ranking stage, providing a practical path to robust LTV learning without re‑ranking infrastructure.

\begin{figure}[t]
  \centering
  \vspace{-5pt}
  \includegraphics[width=1.0\linewidth]{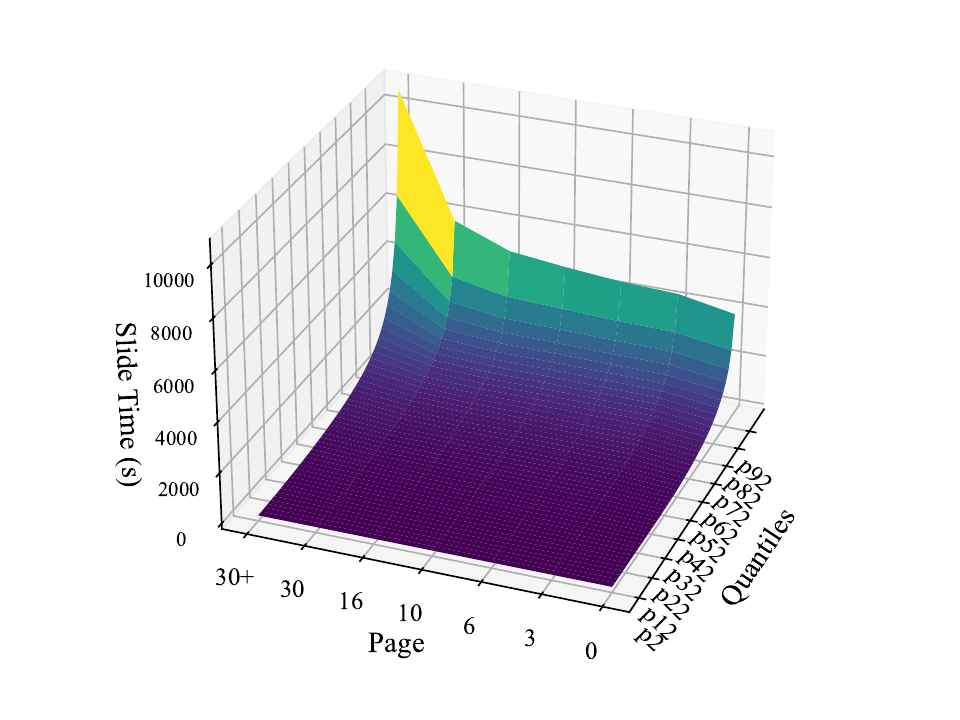}
  \caption{The distribution of slide time corresponding to different quantiles under different page groups.}
  \Description{architecture}
  \label{fig:quantitles} 
\end{figure}

\section{Methods}
In this section, we formalize the LTV optimization problem of the rank stage and describe our methods to predict it. 
The slide time of a card at position $n$ is defined as:
$slide \,time = t_1 + t_2 + ... + t_k$, where $t_k$ represents the video watch time at position $n + k$. Considering the upper limit $Q$ for the influence of the current card on subsequent cards, the final $y_s = Min(t_1 +t_2 + ... + t_k, Q)$. 

\subsection{Position Debias Quantile-based Framework } 
The original slide‑time prediction method is susceptible to inherent position bias, arising from systematic differences in user engagement across exposure positions in recommendation systems. As illustrated in Figure \ref{fig:position_bias}, the video slide time demonstrates a strong correlation with its exposure position. A sharp increase in slide time is observed when the average exposure position surpasses a critical threshold. This phenomenon arises because videos positioned later in the sequence are strategically placed to engage active users who tend to consume more content.

Figure \ref{fig:quantitles} presents the 3D quantile distributions of slide time across pages. Each page returns 4 videos. 
The visualization reveals systematic shifts in quantile patterns across increasing page numbers, with higher page numbers exhibiting elevated slide time values for equivalent quantiles. Color intensity encodes the relative magnitude of slide time, providing visual evidence of position bias.

This motivates our PDQ framework, which models slide-time rankings within page groups via quantile regression. Our framework enables the identification of videos demonstrating superior performance relative to peers within identical exposure contexts while preserving positionally informative behavioral patterns.  The complete training pipeline is outlined in Algorithm 1. 
 The core mechanism partitions the data into $M$ page groups by isofrequency division, enabling group-specific slide time estimation through quantile analysis. For each group $k$, isofrequency quantiles $\{D_{kj}\}_{j=1}^{T}$ are calculated based on slide time distributions. 
 For each page group $ k $, we define $ T $-quantile thresholds $ \{D_{k,j}\}_{j=1}^T $ such that:
\begin{equation}
D_{k,j} = F_k^{-1}\left( \frac{j}{T} \right)
\end{equation}
where $ F_k^{-1}(\cdot) $ is the inverse cumulative distribution function (CDF) of slide times in group $k$. 
 Each sample's quantile label $y_{ki}$ is determined through this analysis. 
 Model training proceeds via minimization of mean squared error (MSE) loss $L(\theta)$. 
$\theta$ represents model parameters, $u_i$ and $v_i$ denote user and video features respectively.

\begin{algorithm}[htbp]
\caption{Training of PDQ: Position Debias Quantile-based Slide time Prediction}
\begin{algorithmic}[1]
\renewcommand{\algorithmicrequire}{\textbf{Input:} Training data $\{(u_i, v_i, p_i, s_i)\}_{i=1}^n$}
\renewcommand{\algorithmicensure}{\textbf{Output:} Slide-time quantile prediction model $\theta$}
\REQUIRE 
\STATE Partition data into page groups: $\{P_k\}_{k=1}^{M}$ based on $\{p_i\}_{i=1}^{n}$
\STATE Calculate isofrequency quantiles $\{D_{kj}\}_{j=1}^{T}$ for each $P_k$
\STATE Generate quantile labels: $y_{ki} \leftarrow \text{Quantile}(s_i, D_{i*})$
\STATE Train model: $\theta^* \leftarrow \arg\min_{\theta} \sum_{i=1}^n \text{MSE}(\theta(u_i,v_i), y_{ki})$
\ENSURE Trained model $\theta^*$
\label{algo:1}
\end{algorithmic}
\end{algorithm}

A critical adjustment addresses zero-valued slide times in lower quantiles. Specifically, for page group 0, slide times below the 32nd percentile are censored at zero due to user interaction patterns.  The darkest part in Figure 3 represents a value of 0. 
We introduce the dynamic starting quantile index $\{S_{k} \}_{k=1}^{M}$ and the bucketized values $B(s_i,D_{i*})$ to handle this. 
Given isofrequency quantiles $ D_{i*} = \{D_{k,j}\}_{j=1}^T $ for the group of pages $ k $, the bucket-based value $ B(s_i,D_{i*}) $ identifies the smallest quantile index containing $ s_i $ and $S_k$ represents the maximum index without zero value: 

\begin{equation}
  y_{ki} = \text{Quantile}(s_i, D_{i*}) =
\frac{1}{T} * (B(s_i,D_{i*}) +S_{k})
\end{equation}

\begin{equation}
B(s_i,D_{i*}) = 
\begin{cases}
\min \left\{ j \in \{1,\dots,T_k\} \,\middle|\, s_i > D_{k,j} \right\}, & s_i > D_{k,1} \\
0, & \text{otherwise}
\end{cases}
\end{equation}

\begin{equation}
S_{k} = \max \left\{ j \in \{1,\dots,T_k\} \,\middle|\, D_{k,j} =0 \right\}
  \end{equation}
This ensures discrete stratification of continuous slide time values into $T$ quantile bins while explicitly handling zero-value samples.
The stratified approach ensures group-specific quantile estimation, thereby improving the precision of position-aware value prediction.
During inference, the PDQ model processes user-video pairs $(u_0,v_0)$ by first identifying their page group $P_k$, then predicting the slide time quantile $\hat{y}_{ki}=\theta(u_i,v_i)$. 
The framework systematically neutralizes positional bias by leveraging within-group ranking structures. Integrating positional information into the label generation process provides stronger supervisory signals than treating position as a standard input feature. 
Notably, this approach maintains constant model complexity and inference latency compared to baseline models while transforming continuous value prediction into a bounded quantile estimation task (0-1 range), effectively reducing estimation difficulty.
The quantization granularity parameter $T=50$ is determined through systematic experimentation. 

\begin{figure}[tbp]
  \centering
  \includegraphics[width=1.0\linewidth]{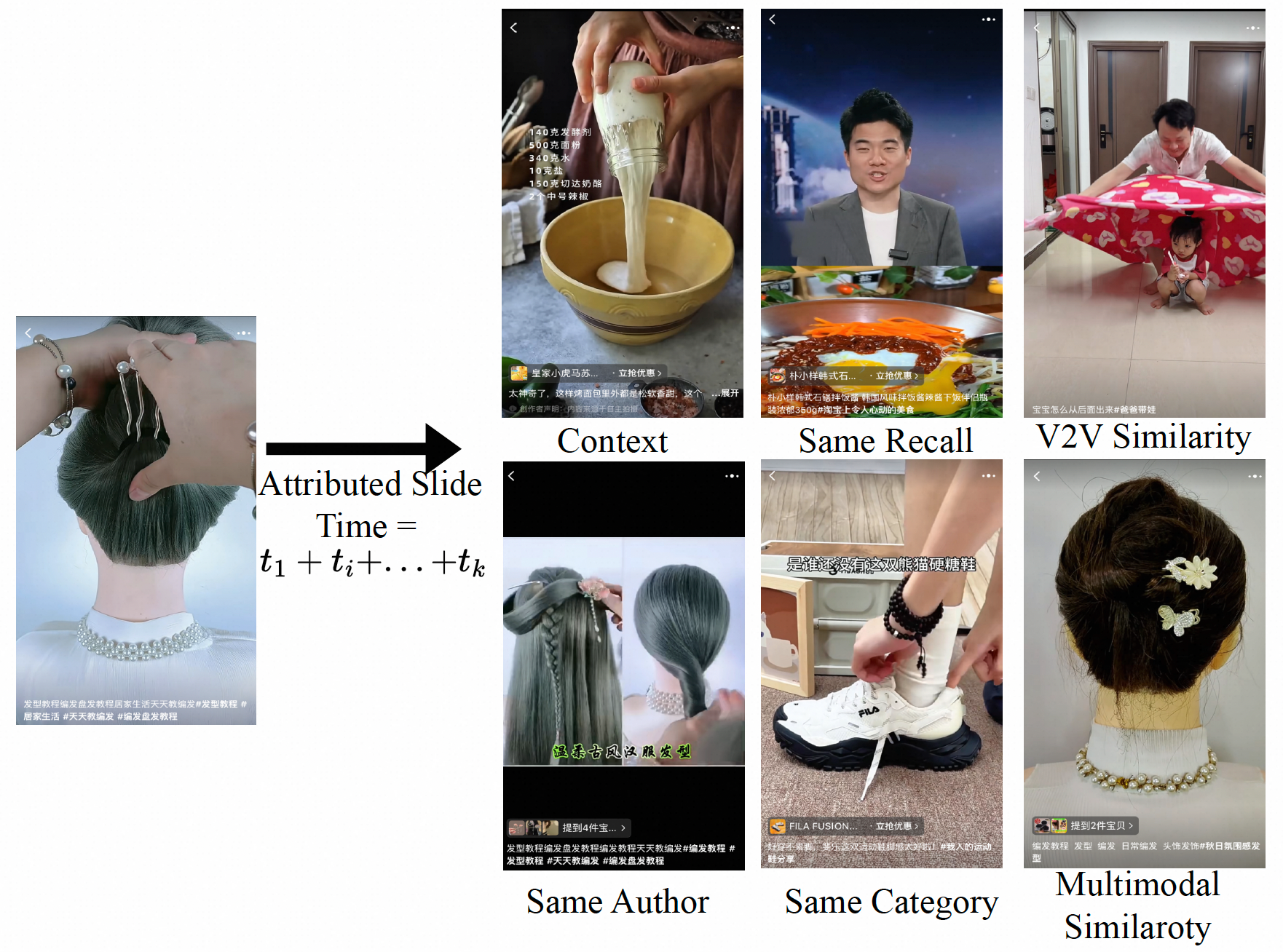}
  \caption{The instance of different value attribution types.}
  \Description{architecture}
  \label{fig.4} 
\end{figure}
\subsection{Long-term Value Attribution}
Cumulative playtime aggregation for LTV measurement has critical limitations: it ignores contextual dependencies and generates noise from spurious correlations between sequential but unrelated content. To address this, we propose multi-dimensional attribution analysis through three dimensions:
contextual dependency, behavioral similarity, and content affinity. 


Let $t_i$ denote the playtime of video $i$. The baseline
slide time $S_j$ for video $j$ is defined as:

\begin{equation}
S_j = \sum_{i=j+1}^n t_i
\end{equation}
Our framework incorporates five fine-grained dimensions (Table \ref{tab:1}) to compute the attributed slide time:
\begin{equation}
S_j = \sum_{i=j+1}^n c_{ji} t_i
\end{equation}

\begin{table}[tbp]
    \centering
    \caption{Multi-Dimensional Attribution Analysis}
    \label{tab:1}
    \begin{tabular}{c|c|cc}
        \hline
        \hline
        Dimension & Type & $S_{ratio \%}$ & $V_{ratio \%}$ \\ \hline
        \hline
        Contextual  & $c_{ji}^{(pos)}=1$ & 34.29 & 11.39 \\
        Dependency & $sgn(c_{ji}^{(pos)}+c_{ji}^{(col)})=1$ & 53.50 & 12.25 \\ \hline
        Behavioral & $c_{ji}^{(rec)}=1$ & 35.04 & 23.42 \\
        Similarity & $c_{ji}^{(v2v)}=1$ & 3.25 & 0.63 \\ \hline
        Content & $c_{ji}^{(mm)}=1$ & 6.33 & 0.91 \\    
        Affinity & $c_{ji}^{(auth)}=1$ & 15.45 & 1.87 \\
         & $c_{ji}^{(cat)}=1$ & 41.99 & 20.21 \\ \hline
    \end{tabular}
\end{table}
Here, $c_i \in[0,1]$ quantifies the causal relationship between the $j$-th and $i$-th videos, reflecting the strength of the correlation.
Table \ref{tab:1} quantifies attribution effectiveness across three dimensions using two metrics: (1) Slide Time Contribution Ratio ($S_{ratio}$) and (2) Video Coverage Ratio ($V_{ratio}$). 
\begin{itemize}
 \item \textbf{Contextual Dependency:} Models sequential patterns through adjacent position co-occurrence $c_{ji}^{(pos)}$ and collection-based transitions $c_{ji}^{(col)}$. 
 
 \item \textbf{Behavioral Similarity:} Captures retrieval consistency $c_{ji}^{(rec)}$ and video-to-video similarity $c_{ji}^{(v2v)}$ using V2V tables. 
 \item \textbf{Content Affinity:} Quantifies cross-modal alignment $c_{ji}^{(mm)}$ via multimodal embeddings, author associations $c_{ji}^{(auth)}$, and category coherence $c_{ji}^{(cat)}$. 
 
 \end{itemize}
 The attribution weights $c_{ji}$ are computed through:
 \begin{dmath}
     c_{ji} = \sigma\left(w^{(pos)}c_{ji}^{(pos)} + w^{(col)}c_{ji}^{(col)} + w^{(rec)}c_{ji}^{(rec)} + w^{(v2v)}c_{ji}^{(v2v)} +  w^{(mm)}c_{ji}^{(mm)}+ w^{(auth)}c_{ji}^{(auth)}+ w^{(cat)}c_{ji}^{(cat)}\right)
 \end{dmath}
 where $\sigma$ denotes the sigmoid function and $\{w^{(d)}\}$ are learnable parameters.
 The correlation coefficients were reduced to binary values of 0 and 1 to simplify the modeling process: 
 \begin{dmath}
     c_{ji} = sgn\left(c_{ji}^{(pos)} + c_{ji}^{(col)} + c_{ji}^{(rec)} + c_{ji}^{(v2v)} +  c_{ji}^{(mm)}+ c_{ji}^{(auth)}+ c_{ji}^{(cat)}\right)
 \end{dmath}
 
As each request yields four items, we integrate the six adjacent position videos ($c_{ji}^{(pos)}$) generated from the current video, along with their associated collection videos ($c_{ji}^{(col)}$), into the slide time. 
$c_{ji}^{(col)}$ linked content with the same retrieval within the same session.
$c_{ji}^{(v2v)}$: Within the same session, we perform pairwise comparisons of videos and attribute those with a V2V similarity score exceeding 0.5.
$c_{ji}^{(mm)}$: Within the same session, we calculate the multimodal similarity between each pair of videos. Videos with a similarity score exceeding 0.9 are attributed accordingly. The multimodal embeddings are extracted with pretrained models CLIP\cite{radford2021learning}, Vindlu\cite{cheng2023vindlu}, and BEIT3\cite{wang2023image}. 
 $c_{ji}^{(auth)}$: Despite low co-occurrence frequency within sessions, author-related videos account for substantial watch time, suggesting strong user engagement with preferred creators.

\begin{figure}[tbp]
  \centering
  \includegraphics[width=1.0\linewidth]{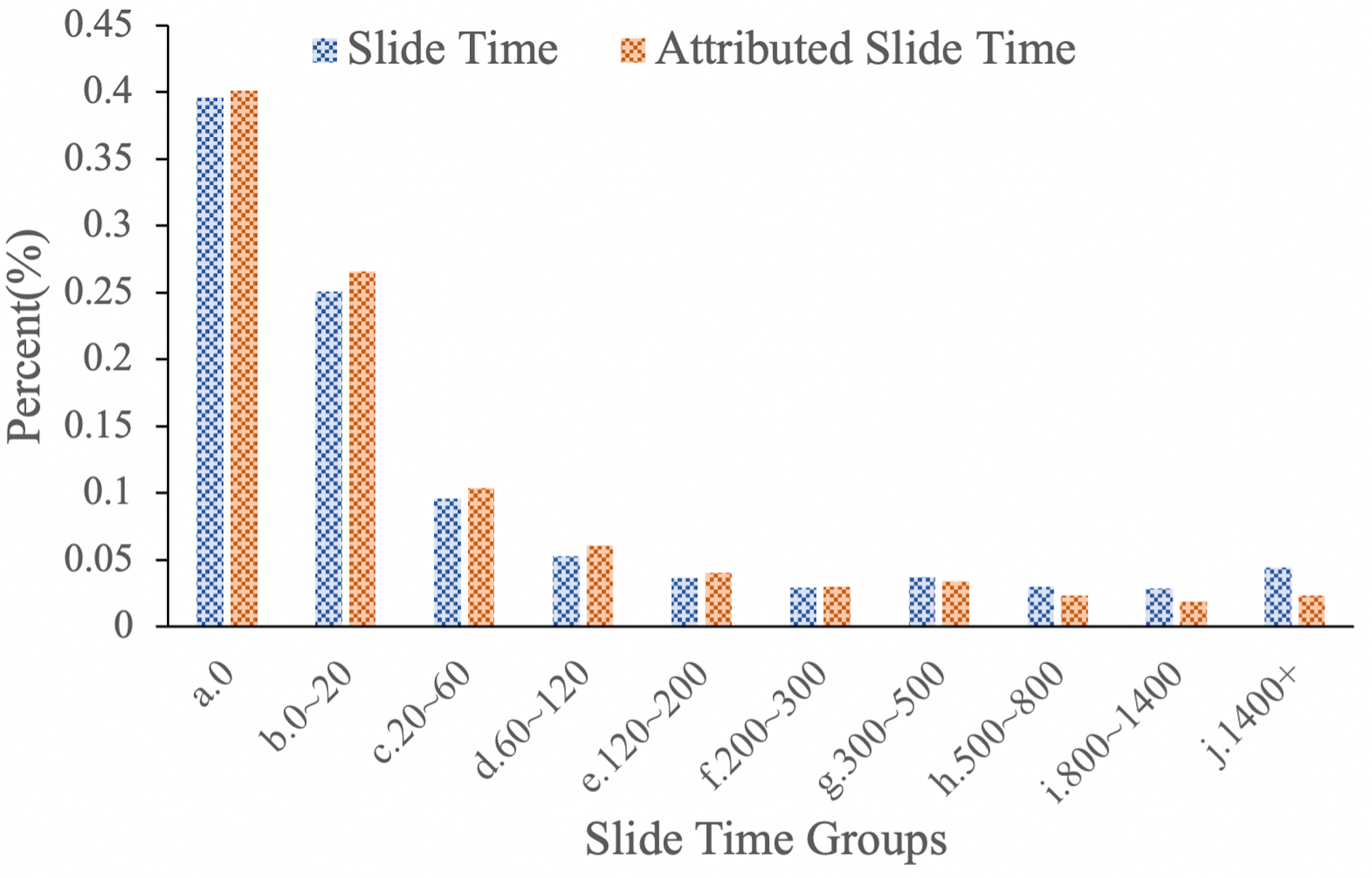}
  \caption{Percentage Distribution of Slide Time and Attributed Slide Time.}
  \Description{architecture}
  \label{fig.slide_time_groups} 
\end{figure}

To address the heavy-tailed nature of $S_j$ as figure \ref{fig.slide_time_groups} shows, we employ compound Poisson-Gamma regression with Tweedie loss\cite{tweedie1984index}:
\begin{equation}
   \mathcal{L}_{Tweedie} = \frac{1}{N} \sum_{i=1}^{N} \left( - y_i\frac{ \mu_i^{1-\rho}}{1-\rho} + \frac{\mu_i^{2-\rho}}{2-\rho} \right) \quad 
   \label{eq:tweedie}
\end{equation}
\begin{equation}
\mathcal{L} = \mathcal{L}_{MSE} + \lambda \mathcal{L}_{Tweedie}
 \label{eq:attribute_loss}
\end{equation}
This formulation naturally handles zero-inflated continuous targets through parameter $\rho$, where $\rho=1.5$ achieves optimal performance in our experiments. The final objective combines mean squared error with Tweedie loss, where $\lambda$ balances the two components.


\subsection{Long-term Value Cycle Modeling}
In the field of short videos, high-quality creators have a huge impact on users. 
According to our analyses, users frequently revisit their preferred authors, often returning over several days.
Author-related interactions occur beyond the immediate session, necessitating multi-day tracking.
Therefore, to more accurately capture users' preferences and interests, we should adopt a broader temporal perspective and conduct in-depth modeling and analysis of the authors that users favor. Specifically, by extending the observation period to seven days, we can more comprehensively assess each author's long-term appeal and potential value to readers. 
These findings motivate our definition of author-centric LTV:  
\begin{equation}
  S^{(t)}_{\text{auth}} = \sum_{d=t-N+1}^{t} \sum_{v \in V_{\text{auth}}} \alpha^{t-d} \cdot t_v(d)
\end{equation}

where $S^{(d)}_{auth}$ quantifies the author's influence over $N$-day window. 
$\alpha$ models the temporal decay of user attention, while $V_{\text{auth}}$ denotes videos by the same author.  $N=7$ strikes an optimal balance between system latency and predictive efficacy, accounting for approximately 6.3\% of the users' total viewing time. User interest typically plateaus after the seventh day, and extending the window beyond this period could result in increased training latency without substantial gains.

The deployment of long-term value prediction introduces two critical system-level challenges: (1) Label delay due to multi-day aggregation, where complete LTV labels require N days to materialize; (2) Temporal misalignment between real-time inference and delayed sample updates. 
To address these challenges, we develop a co-training framework with dual-stream sampling.
We construct daily-level value samples through author-centric aggregation. 
Algorithm \ref {algo:2} implements a dual-stream sampling framework. By synchronizing real-time samples $\mathcal{D}_t$ with delayed author-LTV samples $\mathcal{D}_{t-N}$, the co-training strategy achieves system efficiency and temporal consistency. 
The multiple tasks share the underlying embeddings. When training the author LTV task, we apply a stop gradient to prevent it from affecting the updates of the shared embeddings. For other tasks, the shared embeddings are updated normally.

This method enables us to track both immediate engagement ($t=d$) and delayed effects ($t>d$) through parallel data streams, explicitly capturing delayed user responses through multi-day aggregation.
Besides, the author-level aggregation breaks video-level lifecycle constraints via cross-video coherence and introduces exponential decay $\alpha$ to prioritize recent engagement.


\begin{algorithm}[htbp] 
\caption{Dual-stream Sampling Framework} 
\begin{algorithmic}[1] 
\REQUIRE Daily data stream $\{ \mathcal{D}_1, \mathcal{D}_2, ..., \mathcal{D}_t\}$, cycle length $N$
\FOR {each training day $t$} 
    \STATE Obtain standard sample $\mathcal{D}_t$ and delayed sample $\mathcal{D}_{t-N}$
    \STATE Calculate author-level label $S_{auth}^{(t)}$ and video-level label $S_j$
    \STATE co-training strategy: update shared parameters using $\mathcal{D}_t$ and frozen author representations via $\mathcal{D}_{t-N}$
\ENDFOR 
\ENSURE  Update parameters of LTV task and immediate tasks
\end{algorithmic} 
\label{algo:2}
\end{algorithm}

\begin{figure*}[h]
  \centering
  \includegraphics[width=1\textwidth]{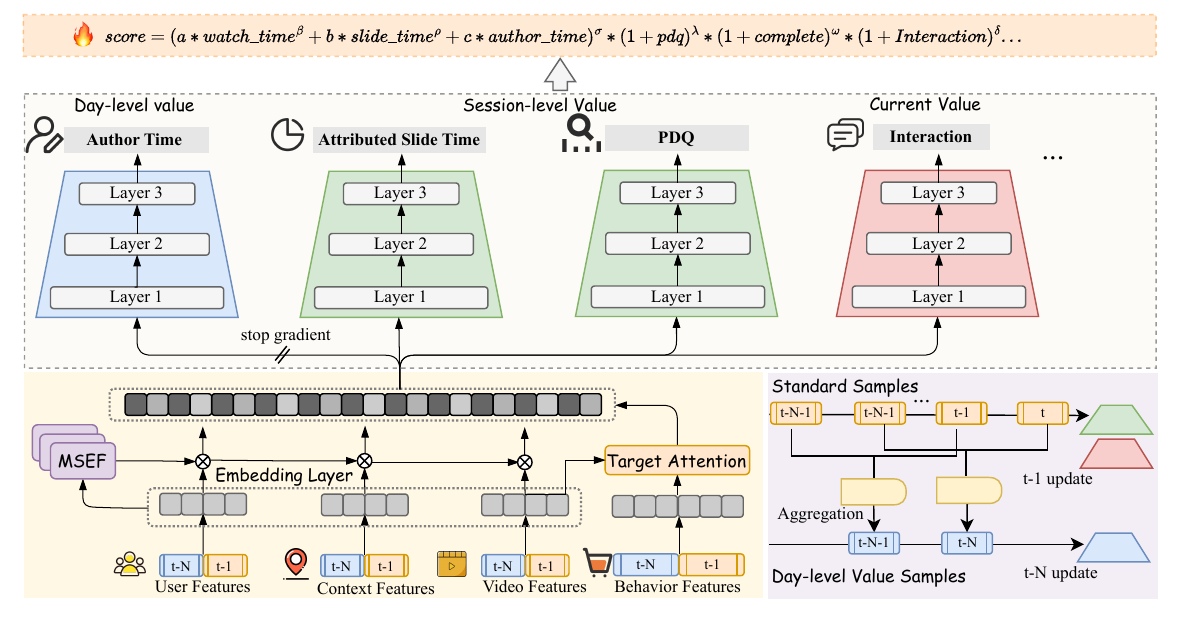}
  \caption{The Architecture of LTV ranking model. The section in blue is updated using daily-level value samples, while the remaining parts are updated using standard samples.  In the online formula, we fused the watch time, attributed slide time, and author time weighted, and then multiplied by the scores and orders of the other targets as the final score.}
  \Description{ltv architecture}
  \label{fig_test_ltv} 
\end{figure*}



\section{Model Framework}
Figure~\ref{fig_test_ltv} summarizes our LTV-oriented ranking framework. The model is trained with two complementary sample streams: standard samples and day-level value samples. We adopt an alternating update schedule— the blue subgraph in Figure~\ref{fig_test_ltv} is optimized with day-level value samples (t–N update), while the remaining parts are optimized with standard samples (t–1 update). Aggregation modules connect the two sample streams, and the stop gradient is applied where indicated to prevent information leakage and stabilize optimization.
Inputs comprise different types of features. User features include profile signals such as user ID, predicted age, and gender. Video features cover video ID, category, and historical statistics. Context features describe environment factors such as location and device model. Behavior features encode historical interactions (e.g., watched,  clicked, carted items) organized as sequences.
At the bottom of the architecture, a Multi-Scale Embedding Fusion (MSEF) layer personalizes the embedding space. MSEF combines a multi-scale feature pyramid (SENet) \cite{hu2018squeeze} that fuses multi-resolution representations to capture semantics at different granularities to emphasize informative dimensions. This yields a shared, target-agnostic representation enriched by cross-dimensional feature interactions.
To capture users’ global interests, we employ a Target Attention module that attends from the target query to the fused multi-source features. 
On top of the shared representation, we build multiple task-specific towers to model heterogeneous objectives: day-level value (e.g., Author Time), session-level value (e.g., Attributed Slide Time, PDQ), and current value (e.g., Interaction, Watch Time and Video Completion Rate). To alleviate optimization conflicts among tasks, we also adopt a PPNet-style multi-objective design~\cite{chang2023pepnet}. The day-level branches are further isolated with stop-gradient, ensuring stable training of long-horizon signals.
At serving, we fuse targets consistent with the formula atop Figure~\ref{fig_test_ltv}: a weighted sum of Watch Time, Attributed Slide Time, and Author Time is multiplicatively calibrated with other heads (PDQ, Completion, Interaction). Weights are tuned offline and verified online to balance short-term engagement and long-term value.
Training uses batch size 512 and AdagradDecay. As the framework is realized as task augmentation on an existing ranker, both warm-up and online deployment require minimal engineering changes and remain compatible with production latency.

\section{Experiments}
We conduct comprehensive offline and online evaluations on an industrial-scale video recommendation platform with 23 million users and 22 million videos to validate the LTV framework.
\subsection{Experimental Setup}
 \textbf{Dataset.}  We use 15 consecutive days of traffic logs from a video recommender system for offline experiments. The first 14 days serve as training data and the final day as test data. Table \ref{table_dataset} summarizes key statistics:

\begin{table}[htbp]
\caption{Overview of the Experimental Dataset}  

\label{table_dataset}  
\resizebox{0.9\linewidth}{!}{
\begin{tabular}{cccc}
\toprule
Dataset       & \#Users & \#Videos & \#Instances   \\ \midrule
 Training         & 23M    & 22M     & 7G          \\ 
 Testing          & 23M    &5M     &523M \\
 \bottomrule
\end{tabular}
}
\label{tab_dataset}
\end{table}



\textbf{Metric.}
We evaluate both session-level (slide time) and day-level (author time) objectives.

\textbf{Offline.}
We report Mean Squared Error (MSE) and Mean Absolute Error (MAE) for point-wise accuracy, XAUC for order consistency, and Predict Click Over Click (PCOC) for calibration.
MSE and MAE quantify the discrepancy between predictions and labels.
XAUC \cite{zhan2022deconfounding}(Eq.~\ref{eq.xauc}) measures how well the ranking induced by predictions agrees with the ranking induced by ground truth; values closer to 1 indicate stronger agreement.
PCOC computes, within calibration buckets, the ratio between the aggregated predicted click propensity and the observed click rate; values closer to 1 indicate better calibration.


\begin{equation}
  \text{XAUC} = \frac{\sum_{i,j} \delta(\hat{y}_i > \hat{y}_j)\delta(y_i > y_j)}{\sum_{i,j} \delta(y_i > y_j)}
\label{eq.xauc}
\end{equation}



\textbf{(2) Online Metric}. We adopt two primary engagement metrics for real-time evaluation: WatchTime (total user viewing duration) and Video Views (VV, total view counts). These metrics reflect immediate user engagement with recommended content.  
For long-term evaluation, we introduce the N-day return visit metric  \( LT_N \), which quantifies sustained user interest:
\begin{equation}
    LT_N = \frac{\sum_{d=1}^{N} \sum_{u \in U} \phi(u)}{|U|}
\end{equation}
\( U \) represents the set of users, and $|U|$ reflects the total number of users in this group. The function \( \phi(u) \) outputs 1 if user \( u \) revisits within \( N \) days, otherwise it outputs 0.
Immediate and long-term metrics help to reflect user interest and long-term engagement.

\subsection{Offline Evaluation}
We individually assessed three proposed methods. In Table \ref{tab_pdq_result}, the baseline refers to the original slide time MSE method.

\subsubsection{PDQ Method}
The PDQ method achieves an XAUC score of 0.6378, marking an improvement of 0.0126 over the baseline score of 0.6252. This suggests that PDQ enhances the predictive accuracy for slide time order. Furthermore, PDQ demonstrates superior calibration performance, reducing MSE from 4.9847 to 0.0946. This quantile-based formulation (0-1 normalized labels) inherently improves prediction stability compared to the original regression targets. 
The XAUC-2 is calculated based on grouped quantiles of slide time, and while this metric is unavailable for the baseline, PDQ achieves a score of 0.6894, surpassing its own XAUC score.
Overall, the PDQ method demonstrates enhanced performance compared to the baseline across most metrics.
\begin{table}[htbp]
\caption{Offiline Evaluation Results of PDQ. Overall Performance between PDQ and Baseline Slide Time.}
\label{pdq_result}  
\setlength{\tabcolsep}{1.1mm}
\begin{tabular}{ccccccc}
\toprule
\textbf{Method}      &           \textbf{XAUC}   & \textbf{MSE}  & \textbf{MAE} &\textbf{PCOC}  & \textbf{XAUC-2}  \\ \midrule
 Slide Time &      0.6252  &  4.9847   & 1.7434   &  0.8485 &  -        \\ 
 PDQ     &       0.6378\textbf{(+0.0126)}   &  0.0946   &0.2402  &   0.8481 & 0.6894       \\
 \bottomrule
\end{tabular}
\label{tab_pdq_result}
\end{table}

\begin{table*}[htbp]
\centering

\caption{Performance Metrics Comparison of Attributed Slide Time.}
\resizebox{\linewidth}{!}{
\begin{tabular}{lccccccc}
\toprule
\textbf{Method}& \textbf{Loss} & \textbf{$r$} & \textbf{$\rho$} &  \textbf{MSE} &  \textbf{MAE} & \textbf{XAUC} & \textbf{PCOC} \\
\midrule
Slide Time &MSE& - & - & 4.9847 & 1.7434 & 0.6253 & 0.8485 \\ 
Attributed ST&MSE & - & - & 4.1092(-0.8755) & 1.5340(-0.2094) & 0.6371(+0.0118) & 0.8809(+0.0324) \\
Attributed ST&Tweedie & 1.0 & 1.5 & \textbf{3.7971(-1.1876)} & \textbf{1.4527(-0.2907)} & 0.6355(+0.0102) & \textbf{0.9184(+0.0699)} \\
Attributed ST&Tweedie+MSE & 0.1 & 1.5 & 4.0867(-0.8980) & 1.5323(-0.2111) & 0.6368(+0.0115) & 0.8819(+0.0334) \\
Attributed ST&Tweedie+MSE & 0.1 & 1.4 & 4.3176(-0.6671) & 1.5941(-0.1493) & 0.6360(+0.0107) & 0.8580(+0.0095) \\
Attributed ST&Tweedie+MSE & 0.1 & 1.3 & 4.0953(-0.8894) & 1.5334(-0.2100) & \textbf{0.6377(+0.0124)} & 0.8816(+0.0331) \\
\bottomrule
\label{tab:attribute_results}
\end{tabular}
}
\end{table*}

\begin{table}[htbp]
    \centering
    \caption{PDQ Offline Results During Different Page Groups.}
    \begin{tabular}{l|c|c|l}
        \toprule
        \textbf{Page Groups} & \textbf{Pair Count} & \textbf{Base XAUC} & \textbf{PDQ XAUC}  \\
        \midrule
        a.0 & \( 6.5\times 10^{8} \) & 0.4917 & 0.3936  \\
        b.1-2 & \( 6.4\times 10^{8} \) & 0.7230 & 0.7581 \textbf{(+0.0351)}  \\
        c.3-5 & \( 9.2\times 10^{8} \) & 0.8037 & 0.8378 \textbf{(+0.1341)}  \\
        d.6-9 & \( 1.1\times 10^{9} \) & 0.8012 & 0.8010  \\
        e.10-15 & \( 1.4\times 10^{9} \) & 0.7748 & 0.8425 \textbf{(+0.0677)}  \\
        f.16-29 & \( 2.3\times 10^{9} \) & 0.7175 & 0.8855 \textbf{(+0.1680)}  \\
        m.30+ & \( 8.3\times 10^{9} \) & 0.6502 & 0.6887 \textbf{(+0.0385)}  \\
        \bottomrule
    \end{tabular}
    \label{tab:pdq_page_groups}
\end{table}

We conducted an in-depth analysis of the PDQ offline metrics in Table \ref{tab:pdq_page_groups}. 
Since the slide time of the video in the top position of the same session will be higher than that of the video in the bottom position during the calculation of XAUC, we limit the comparison under the same page group to ensure the fairness of the metrics.
Pair count represents the number of pairs of samples in each page group. The evaluation is performed by grouping data according to pages. Except for requests on page 0, there is generally an improvement in the XAUC for other page requests. This demonstrates that the task can effectively neutralize position bias and enhance the accuracy of slide time estimates. 
The anomaly on page 0 is attributed to external factors such as external distributions, where some content is promoted through external channels.


\subsubsection{Attributed Method}
In our experiments, we performed a series of ablation studies to assess the effectiveness of our method on both the original and attributed slide time, as detailed in Table \ref{tab:attribute_results}. The parameters \( r \) and \( \rho \) denote the weight of the Tweedie loss in Eq. \ref{eq:attribute_loss} and its parameter in Eq.\ref{eq:tweedie}, respectively. 
For the model using only attribute labe+mse, we observed an improvement in XAUC from 0.6253 to 0.6371, an increase of 0.0118, while the MSE decreased significantly by 0.8755. 
By integrating the Tweedie loss optimization, performance is further enhanced, with the MSE decreasing from 4.1092 to 3.7971. This adjustment effectively mitigated the underestimation issue, resulting in a more precise score. 


\begin{table}[htbp]
    \centering
    \caption{Online Results of Proposed Methods.}
    \resizebox{\linewidth}{!}{
    \begin{tabular}{l|cccccc}
        \toprule
       \textbf{Task} & \textbf{VV} & \makecell{\textbf{Watch} \\ \textbf{Time}} & $\boldsymbol{LT_1}$ & $\boldsymbol{LT_3}$ & \makecell{\textbf{QA} \\ \textbf{VV}} & \makecell{\textbf{QA} \\ \textbf{Watch Time}} \\
        \midrule
        PDQ & \underline{2.49}\% & -0.04\% & 0.00\% & 0.02\% & - & - \\
        Attributed ST & -1.92\% & \underline{1.23\%} & 0.08\% & 0.11\% & - & - \\  
        Author Time & -0.50\% & 0.35\% & 0.16\% & \underline{0.21\%} & \underline{4.03\%} & 2.60\% \\
        \bottomrule
    \end{tabular}
    }
    \label{tab:online_metrics}
\end{table}

\subsubsection{Author Time Method}
We evaluate two training strategies for the Author Time task aligned with our cross‑temporal design.
(1) Single model: start a dedicated branch warm‑started from the production ranker, updated only on t–N day‑level samples to prevent interference with other tasks.
(2) Co‑training: train Author Time together with session/current‑value heads in the unified ranker under the alternating t–1/t–N schedule, sharing the backbone and Target Attention. 
Figure \ref{fig.authortime_results} reveals that the co-training method offers additional improvements over using a single model on PCOC, potentially benefiting from the supportive role of other tasks. 
These results suggest that the co-training method provides complementary contextual/behavioral/content signals that regularize the shared representation, and the stop-gradient plus alternating updates stabilize learning under delayed labels.

\begin{figure}[tbp]
  \centering
  \includegraphics[width=1.0\linewidth]{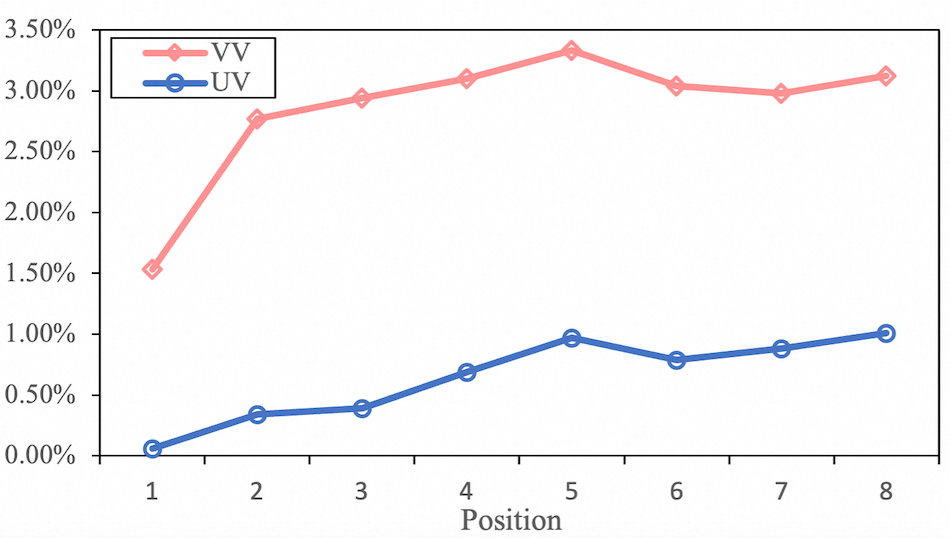}
  \caption{UV and VV improvement at different positions.}
  \Description{day_level_sample}
  \label{fig.pdq_online_results} 
  \vspace{-5pt}
\end{figure}
\begin{figure}[tbp]
  \centering
  \includegraphics[width=\linewidth]{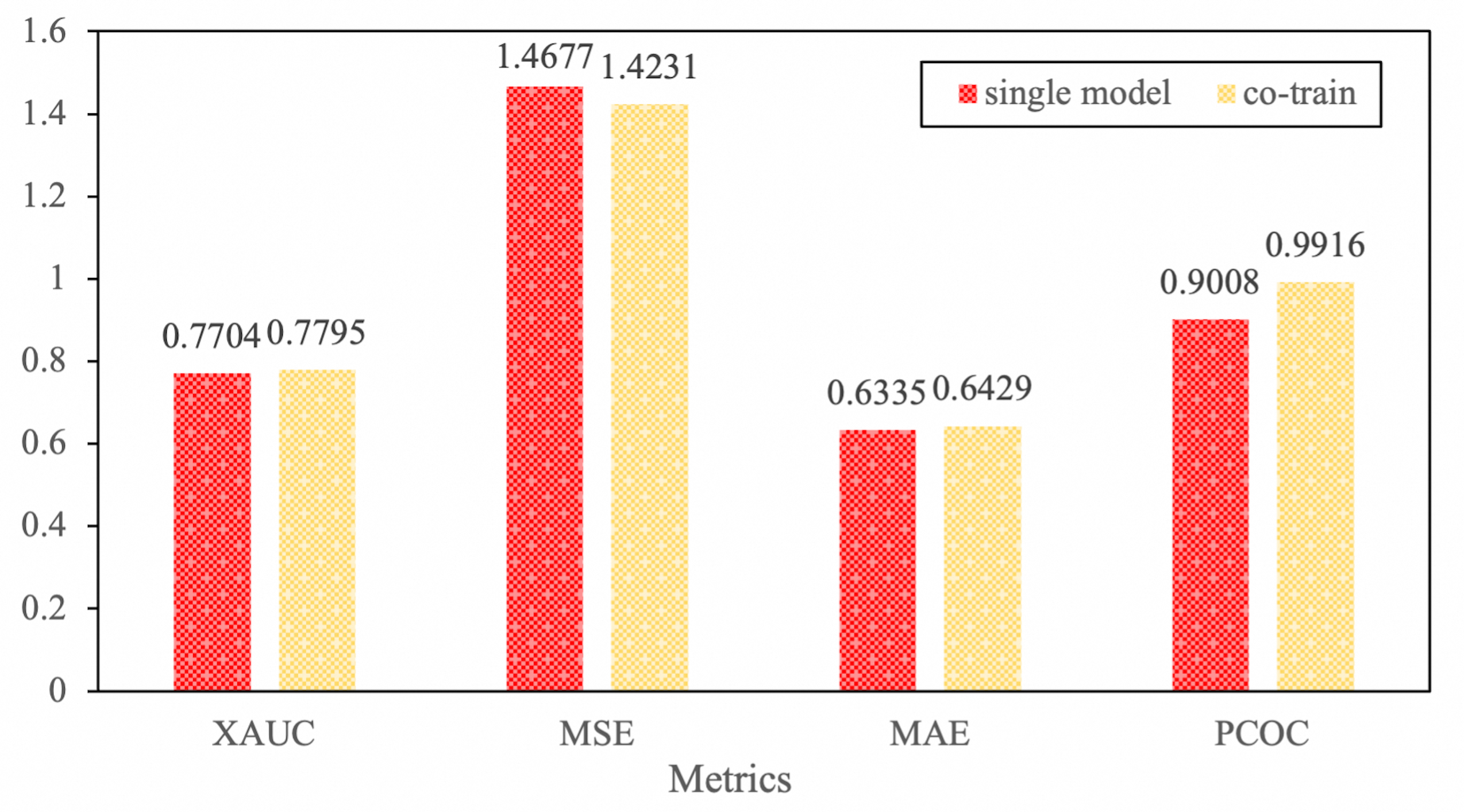}
  \caption{Metrics of author time with different training methods.}
  \Description{day_level_sample}
  \label{fig.authortime_results} 
\end{figure}


\subsection{Online Experiments}
We integrate the LTV modules into the Taobao App’s production ranker via a unified multi‑objective framework. 
We integrate the LTV modules into the Taobao App’s production ranker via a unified multi‑objective framework that preserves both raw scores and the score‑induced order.
We report relative lifts versus the baseline; all changes with underline are statistically significant at conventional levels over the multi‑day horizon.

Table \ref{tab:online_metrics} presents the performance results of the methods evaluated with online testing over several days on the Taobao App. 
We use the MSE method for the initial slide time as our baseline, which has already gained an improvement of 4\% VV. 
The results show the great success of our proposed long-term framework. Among them, PDQ achieves a larger improvement of $2.49\%$ for VV. Figure \ref{fig.pdq_online_results} shows the improvement in the slide rate of different positions of the PDQ target, and the increase of UV and VV can demonstrate the increase in the slide rate of users.
QA VV and QA watch time refer to the VV and time spent on high-quality authors, who are vital contributors to the content ecosystem. 
Author LTV achieves statistically significant gains in long-term retention ($LT_3: +0.21\%$) and quality author engagement (QA VV: +4.03\%).
The attribute slide time task exhibits a gain of $1.23\%$ on watch time at the expense of $1.92\%$ VV, representing a cost-effective approach. 
The three targets each played a distinct role. 
PDQ mitigates position bias; Attributed Slide Time improves intra‑session attribution; Author Time strengthens cross‑day signals for high‑quality creators. 
In all tests, system guardrails (latency, error rates) remained within production thresholds, and the unified fusion preserved ranking stability during traffic ramps.

\section{Conclusion}


We propose a practical ranking-stage framework for long-term value (LTV) that bridges the gap between immediate engagement optimization and long-term value maximization in short‑video recommendation. The design addresses three challenges—position bias (via PDQ quantile normalization without architectural changes), attribution ambiguity (via learnable multi‑dimensional attribution across contextual, behavioral, and content signals with explicit noise filtering through a customized hybrid loss), and limited temporal scope (via cross‑temporal author modeling with day‑level targets)—and fuses long‑ and short‑term signals in a unified ranker. The framework integrates as a lightweight task augmentation, preserving training/serving efficiency and enabling billion‑scale deployment; offline and online results show significant LTV gains with stable trade‑offs against short‑term objectives. Limitations point to clear extensions: broadening cross‑temporal modeling beyond authors to topics/styles and incorporating additional negative signals. We leave these directions, alongside further learnable attribution, to future work.

\bibliographystyle{ACM-Reference-Format}
\balance
\bibliography{main}

\appendix

\end{document}